\documentclass[10pt]{article}
\usepackage{array}
\usepackage{subfig}
\usepackage{booktabs}
\usepackage{amsfonts,amsmath,amssymb,amsbsy,amsfonts, amsthm,epsfig,graphicx,rotating}
\usepackage{latexsym}
\usepackage{bm}
\usepackage{grffile}
\usepackage{ifpdf}
\usepackage{epstopdf}
\usepackage{setspace}
\usepackage{epsf}
\usepackage{bm}
\usepackage{enumerate}
\usepackage{framed}
\usepackage{xcolor}
\usepackage[round,comma,authoryear]{natbib}
\usepackage{amsmath, amssymb, }
\usepackage{epsfig}
\RequirePackage{amsopn}
\usepackage[english]{babel}
\usepackage{algorithm}
\usepackage{algorithmic}
\usepackage{multirow}
\usepackage{todonotes}
\usepackage{color}
\usepackage{hyperref}

\newcommand{\yb}{\bm{y}}
\newcommand{\fb}{\bm{f}}
\newcommand{\xb}{\bm{x}}
\newcommand{\Yb}{\bm{Y}}
\newcommand{\Ybmat}{\mathbf{Y}}
\newcommand{\ob}{\bm{0}}

\newcommand{\sD}{\sf{D}}

\newcommand{\Kb}{\mathbf{K}}
\newcommand{\Mb}{\mathbf{M}}
\newcommand{\Ib}{\mathbf{I}}
\newcommand{\Xb}{\mathbf{X}}
\newcommand{\Fb}{\mathbf{F}}
\newcommand{\Sb}{\mathbf{S}}
\newcommand{\Ab}{\mathbf{A}}
\newcommand{\Bb}{\mathbf{B}}
\newcommand{\Cb}{\mathbf{C}}
\newcommand{\Ncal}{\mathcal{N}} 
 
\newcommand{\Sigmab}{\bm{\Sigma}}
\newcommand{\Omegab}{\bm{\Omega}}
\newcommand{\Omegabhat}{\hat{\bm{\Omega}}}
\newcommand{\Gcal}{\mathcal{G}} 

\newcommand{\tr}{\text{tr}}
\newcommand{\LOOCV}{\text{LOOCV}}

\newcommand{\TP}{\text{TP}}
\newcommand{\TN}{\text{TN}}

\newcommand{\FP}{\text{FP}}
\newcommand{\FN}{\text{FN}}
\newcommand{\KLCV}{\text{KLCV}}

\newcommand{\vect}{\text{vec}}
\newcommand{\KL}{\text{KL}}

\newcommand{\AIC}{\text{AIC}}
\newcommand{\BIC}{\text{BIC}}
\newcommand{\df}{\text{df}}
\newcommand{\bias}{\text{bias}}

\newcommand{\omegahat}{\hat{\omega}}

\newtheorem{lemma}{Lemma}
\begin{document}
\renewcommand{\baselinestretch}{1.2}
\markright{
}
\markboth{\hfill{\footnotesize\rm IVAN VUJA\v{C}I\'C , ANTONINO ABBRUZZO AND ERNST C. WIT
}\hfill}
{\hfill {\footnotesize\rm A computationally fast alternative to cross-validation in penalized Gaussian graphical models} \hfill}
\renewcommand{\thefootnote}{}
$\ $\par
\fontsize{10.95}{14pt plus.8pt minus .6pt}\selectfont
\vspace{0.8pc}
\centerline{\large\bf A computationally fast alternative to cross-validation }
\vspace{2pt}
\centerline{\large\bf in penalized Gaussian graphical models}
\vspace{.4cm}
\centerline{Ivan Vuja\v{c}i\'c$^{\textrm{1}}$, Antonino Abbruzzo$^{\textrm{2}}$ and Ernst C. Wit$^{\textrm{1}}$}
\vspace{.4cm}
\centerline{\it $^{\textrm{1}}$University of Groningen, $^{\textrm{2}}$University of Palermo }
\vspace{.55cm}
\fontsize{9}{11.5pt plus.8pt minus .6pt}\selectfont

\footnotetext{Corresponding author: Ivan Vuja\v{c}i\'c ( i.vujacic@rug.nl) }
\begin{quotation}
\noindent {\it Abstract:}
We study the problem of selection of regularization parameter in penalized Gaussian graphical models. When the goal is to obtain the model with good predicting power, cross validation is the gold standard.
We present a new estimator of Kullback-Leibler loss in Gaussian Graphical model which provides a computationally fast alternative to cross-validation. The estimator is obtained by approximating leave-one-out-cross validation. Our approach is
demonstrated on simulated data sets for various types of graphs. The proposed formula exhibits superior performance, especially in the typical small sample size scenario, compared to other available alternatives to cross validation, such as
Akaike's information criterion and Generalized approximate cross validation. We also show that the estimator can be used to improve the performance of the BIC when the sample size is small.\par

\vspace{7pt}
\par
\end{quotation}\par


\section{Introduction}
\label{sec1}
Let $\Yb=(Y_1,\ldots,Y_p)$ be a $p$-dimensional Gaussian random vector with zero mean and positive definite covariance matrix $\Sigmab$, i.e. $\Yb\sim \Ncal_p(\ob,\Sigmab)$.
In many applications, like gene network reconstruction, estimating the precision matrix, denoted by $\Omegab=(\omega_{ij})=\Sigmab^{-1}$ is of main interest.
The element $\omega_{ij}$ in $\Omegab$ is proportional to the partial correlation between the $i$th and $j$th components of $\Yb$ conditional 
on all others. Consequently $\omega_{ij}=0$ if and only if $Y_i$ and $Y_j$ are conditionally independent given the rest of the variables in $\Yb$. This gives the appealing graphical interpretation
of vector $\Yb$ as a Gaussian graphical model \citep{dempster1972covariance,lauritzen1996graphical,edwards2000introduction,whittaker2009graphical}.  Vector $\Yb$ can be represented by an undirected graph $\Gcal=(V,E)$, where $V$
 is the set of vertices corresponding to the $p$ coordinates of the vector $\Yb$ and the edges $E=(e_{ij})_{1\leq i<j\leq p}$ represent conditional dependency relationships between variables $Y_i$ and $Y_j$.
The edge $e_{ij}$ between $Y_i$ and $Y_j$ exists if and only if $\omega_{ij}\neq0$. Hence, for estimating the graphical structure it is not only important to estimate the parameters but also to identify
the null entries in the precision matrix. 
\newline
A popular method for precision matrix estimation is the penalized likelihood method \citep{yuan2007model,banerjee2008model,friedman2008sparse,fan2009network}. This method is based on the optimization of 
an objective function which is the sum of the scaled likelihood and some penalty function of the precision matrix. Popular penalties are LASSO, SCAD and adaptive LASSO \citep{lam2009sparsistency,fan2009network}. The selection of the
tuning parameter in this method is equivalent with the model selection of a particular graphical model. The methods that have been used in the literature for selecting the regularization parameter include the Bayesian Information 
Criterion (BIC) \citep{yuan2007model,schmidt2010graphical,menendez2010gene,lian2011shrinkage,gao2012tuning}, the Extended Bayesian Information Criterion (EBIC) \citep{foygel2010ebic,gao2012tuning}, Stability Approach to 
Regularization Selection (StARS) \citep{liuroederwasserman10}, Cross-validation (CV) \citep{rothman2008sparse,fan2009network,schmidt2010graphical,fitch2012computationally}, Generalized Approximate Cross Validation (GACV)
\citep{lian2011shrinkage} and the Aikaike's Information Criterion (AIC) \citep{menendez2010gene,liuroederwasserman10,lian2011shrinkage}.
\par
If the aim is graph identification then the criteria BIC, EBIC and StARS are appropriate. BIC is shown to be consistent for penalized graphical models with adaptive LASSO and SCAD penalties for fixed $p$ \citep{lian2011shrinkage,gao2012tuning}. Numerical results suggest that
BIC is not consistent with the LASSO penalty  \citep{foygel2010ebic,gao2012tuning}. When also $p$ tends to infinity EBIC is shown to be consistent for the graphical LASSO, 
though only for decomposable graphical models  \citep{foygel2010ebic}. The disadvantage of EBIC is that it includes an additional parameter that needs to be tuned. \cite{gao2012tuning} fix this parameter to one and show that in this case 
EBIC is consistent with the SCAD penalty. StARS has the property of partial sparsistency which means that
when the sample size goes to infinity all the true edges will be included in the selected model \citep{liuroederwasserman10}. 
\newline
On the other hand, using cross-validation (CV), generalized approximate cross-validation (GACV) and AIC will result with a model with a good predicting power. Cross-validation and AIC are both estimators of the Kullback-Leibler (KL) information
\citep{yanagihara2006bias}, which under some assumptions are asymptotically equivalent \citep{stone1977asymptotic}. GACV is also an estimator of KL since it is  derived as an approximation to
leave-one-out cross-validation (LOOCV) \citep{lian2011shrinkage}. Advantage of AIC and GACV is that they are not as computationally expensive as CV.
\par
In this paper, we propose an estimator of KL of the model defined by the estimated precision matrix. The Kullback-Leibler information or divergence \citep{kullback1951information} is also known as the entropy loss. The formula that
 we propose exhibits superior performance compared to its competitors AIC and GACV. As it is the case with CV, using the proposed estimator will result with the model that has good predictive power. For the graph identification
 problem, we show how our estimator can be used to improve the performance of the BIC when the sample size is small. 
\par
The rest of the paper is organized as follows. In section \ref{sec:VS} we present an example which clarifies the purpose of different selection methods.
In Section \ref{sec:KLCV} a closed-form approximation of leave-one-out-cross validation is proposed and its derivation is given in Section \ref{sec:derivation}. Section \ref{sec:implementation} covers the details of the implementation of the method, while 
Section \ref{sec:sim_study} includes a simulation study that shows the performance of the proposed estimator. Finally, we discuss the usage of the obtained estimator to graph identification problem in Section \ref{sec:graph_estimation}.
We conclude with Section \ref{sec:conclusion}. Appendix contains proofs and auxiliary material.

\section{Prediction power VS graph structure}
\label{sec:VS}

Let $\Omegab_0$ be a precision matrix that corresponds to the true non-complete graph $\Gcal$ and let $\Omegab_{\epsilon}$ be the matrix obtained by adding  
$\epsilon>0$ to every entry of matrix $\Omegab$. The matrix $\Omegab_{\epsilon}$ is positive definite since it is a sum of one positive definite matrix and one positive semi-definite matrix.
Indeed, $\Omegab_{\epsilon}=\Omegab_0+\xb_{\epsilon}{\xb_{\epsilon}}^{\top}$, where $\xb_{\epsilon}=(\sqrt{\epsilon},\ldots,\sqrt{\epsilon})^{\top}$ is a vector of dimension $p$. Hence, $\Omegab_{\epsilon}$ belongs to the class
of precision matrices and it corresponds to some graph $\Gcal_{\epsilon}$. The Kullback-Leibler divergence of $\Ncal_p(\ob,\Omegab_{\epsilon}^{-1})$ from $\Ncal_p(\ob,\Omegab_0^{-1})$, denoted by $\KL(\Omegab_0;\Omegab_{\epsilon})$, is equal to 
\begin{equation}
\label{klloss}
\KL(\Omegab_0;\Omegab_{\epsilon})=\frac12\{ \tr(\Omegab_0^{-1}\Omegab_{\epsilon})-\log|\Omegab_0^{-1}\Omegab_{\epsilon}|-p\}.
\end{equation}
(see \citep{penny2001kullback}).
Since $\epsilon\rightarrow 0$ implies $\Omegab_{\epsilon}\rightarrow \Omegab_0$, by continuity of log determinant and trace it follows that
$$\lim_{\epsilon\downarrow 0}\KL(\Omegab_0;\Omegab_{\epsilon})=0.$$
However, for every $0<\epsilon<\min_{i,j}|\omega_{ij}|$ the matrix $\Omegab_{\epsilon}$ is a matrix without zero entries and consequently the graph $\Gcal_{\epsilon}$ is the full graph.
Thus, the conclusion is that even though a matrix can be close to the precision matrix of the true distribution with respect to KL loss,
the corresponding graph can be completely different from the true one.
\par
Since CV, AIC and GACV are estimators of KL they should be used for obtaining the model with a good predictive power. For graph identification, BIC, EBIC and StARS are more appropriate, because of their graph selection consistency
properties. Consequently, we treat these two problems separately. Next section we devote to a new estimator of KL and in Section \ref{sec:graph_estimation} we show how it can be used to improve the performance of E(BIC).

\section{KLCV: An approximation of leave-one-out-cross validation}
\label{sec:KLCV}
In this section we introduce a closed-form approximation of leave-one-out-cross validation (LOOCV) that we call {\it Kullback-Leibler cross-validation} (KLCV). The reason for this terminology comes from the fact that cross-validating
the log-likelihood loss provides an estimate to Kullback-Leibler divergence \citep{kullback1951information}.  
\newline
Suppose we have $n$ multivariate observations of dimension $p$ from distribution
$\Ncal_p(\ob,\Omegab_0^{-1})$. Using the notation  $\Sb_k=\yb_k{\yb_k}^{\top}$ for the empirical covariance matrix of a single observation, we have that the empirical covariance matrix is given as $\Sb=1/n\sum_{k=1}^n \Sb_k$. The 
log-likelihood of the data is, up to an additive constant, $l(\Omegab)=n\{\log|\Omegab|-\tr(\Omegab \Sb)\}/2$. When $n>p$ the precision matrix $\Omegab=\Sigmab^{-1}$ can estimated by maximizing the scaled log-likelihood function 
$$\frac{2}{n}l(\Omegab)=\log|\Omegab|-\tr(\Omegab \Sb),$$
over positive definite matrices $\Omegab$.  The global maximizer is the maximum likelihood estimator (MLE) given by $\Omegabhat=\Sb^{-1}$. When $n\leq p$ MLE does not exist.  If $n>p$ and the 
true precision  matrix is known to be sparse, the MLE has a non-desirable property: with probability one all elements of the precision matrix are nonzero.  An alternative approach which yields a sparse estimator can be obtained by 
maximizing
\begin{equation}
\label{penloglik}
\Omegabhat_{\lambda}=\mathrm{argmax}_{\Omegab}\log|\Omegab|-\tr(\Omegab \Sb)-\sum_{i=1}^p\sum_{j=1}^pp_{\lambda_{ij} }(|\omega_{ij}|),
\end{equation}
over positive definitive matrices $\Omegab$. Here, $p_{\lambda_{ij} }$ is a penalty function and $\omega_{ij}$ is the $(i,j)$ element of matrix $\Omegab$ and 
$\lambda_{ij}>0$ is the corresponding regularization parameter.
\par
Let the maximum penalized likelihood estimator (MPLE) $\Omegabhat_{\lambda}$ be defined by (\ref{penloglik}) and let $\KL(\Omegab_0;\Omegabhat_{\lambda})$ be the Kullback-Leibler divergence of the model $\Ncal_p(\ob,\Omegabhat_{\lambda}^{-1})$ from the true distribution
 $\Ncal_p(\ob,\Omegab_0^{-1})$. According to (\ref{klloss}) we have that 
$$\KL(\Omegab_0;\Omegabhat_{\lambda})= -\frac{1}{n}l(\Omegabhat_{\lambda})+\bias,$$
where $l(\Omegab)=n\{\log|\Omegab|-\tr(\Omegab \Sb)\}/2$ and $\bias=\tr(\Omegabhat_{\lambda}(\Omegab_0^{-1}-\Sb))/2$.
We propose an estimator of the Kullback-Leibler divergence of the model $\Ncal_p(\ob,\Omegabhat_{\lambda}^{-1})$ to the true distribution 
\begin{equation}
\label{klcv}
\KLCV(\lambda)= -\frac{1}{n}l(\Omegabhat_{\lambda})+\widehat{\bias}_{\KLCV},
\end{equation}
where 
\begin{equation}
\label{bias_klcv}
\widehat{\bias}_{\KLCV}=1/n(n-1)\sum_{k=1}^n\vect\{(\Omegabhat_{\lambda}^{-1}-\Sb_k)\circ \Ib_{\lambda}\}^{\top}\vect[\Omegabhat_{\lambda}\{(\Sb-\Sb_k)\circ \Ib_{\lambda}\}\Omegabhat_{\lambda}]
\end{equation}
and $\Ib_{\lambda}$ is the indicator matrix, whose entry is $1$ if the corresponding entry in the precision matrix $\Omegabhat_{\lambda}$ is nonzero and zero if the corresponding entry in the precision matrix is zero.  Here, $\circ$
is the  Schur or Hadamard product of matrices and $\vect$ is the vectorization operator which transforms a matrix into a column vector obtained by stacking the columns of the matrix on top of one another.
\newline
In this paper we propose to select $\Omegabhat_{\lambda^*}$ for that $\lambda^*$ that minimizes $\KLCV(\lambda)$ over $\lambda>0$. The resulting estimator will give a model with good predictive power.
While for the MLE we do not need any assumptions  to derive the $\KLCV$, for the MPLE the derivation uses the assumption of the sparsistency of the estimator. An estimator is {\it sparsistent}
if all parameters in the true precision matrix that are zero are estimated as zero with probability tending to one when sample size tends to infinity \citep{lam2009sparsistency}.\newline

\section{Derivation of the KLCV}
\label{sec:derivation}
\subsection{Derivation for the MLE}

We follow the idea of \cite{xiang1996generalized}, i.e.  we introduce an approximation for $\LOOCV$ via several first order Taylor expansions. \cite{lian2011shrinkage} uses the idea to derive GACV for MPLE in GGM, where 
in deriving the formula, the partial derivatives corresponding to the zero elements of the precision matrix are ignored. Here, unlike there, we apply the idea only for MLE estimator and therefore we avoid all technical difficulties
that ignoring the derivatives entails. In the next section we extend the derived formula for MLE to MPLE. Denote the log-likelihood of observation $\yb_k$ with 
$$l_k(\Omegab)=\frac12 \left\{\log|\Omegab|-\tr(\Omegab \Sb_k)\right\}$$
and consider the following function of two variables
$$f(\Sb,\Omegab)=\frac{2}{n}l(\Omegab)=\log|\Omegab|-\tr(\Omegab\Sb).$$
With this notation we have the identity
\begin{equation}
\label{eq3:f_identity}
\sum_{k=1}^n f(\Sb_k,\Omegab)=nf(\Sb,\Omegab).
\end{equation}
Let $\Omegabhat^{(-k)}$ be the estimator of the precision matrix defined in (\ref{penloglik}) with $\lambda_{ij}=\lambda=0$ based on the data excluding the $k$th data point. The leave-one-out cross validation score (see \cite{yanagihara2006bias})
is defined by 
\begin{eqnarray*}
 \LOOCV&=&-\frac{1}{n}\sum_{k=1}^n  l_k(\Omegabhat^{(-k)})=-\frac{1}{2n}\sum_{k=1}^n f(\Sb_k,\Omegabhat^{(-k)})\\
&=&-\frac{1}{2n}\sum_{k=1}^n\{ f(\Sb_k,\Omegabhat^{(-k)})- f(\Sb_k,\Omegabhat)+ f(\Sb_k,\Omegabhat)\}\\
&\stackrel{\eqref{eq3:f_identity}}{=}&-\frac{1}{2}f(\Sb,\Omegabhat)-\frac{1}{2n}\sum_{k=1}^n\left\{f(\Sb_k,\Omegabhat^{(-k)})-f(\Sb_k,\Omegabhat)\right\}\\
   &\approx&-\frac{1}{n}l(\Omegabhat)-\frac{1}{2n}\sum_{k=1}^n\left\{\frac{d f(\Sb_k,\Omegabhat)}{d\Omegab}\right\}^{\top}\vect(\Omegabhat^{(-k)}-\Omegabhat).
\end{eqnarray*}

Using matrix differential calculus (see the Appendix) we have $d f(\Sb_k,\Omegabhat)/d\Omegab=\vect(\Omegabhat^{-1}-\Sb_k)^{\top}$. The term $\vect(\Omegabhat^{(-k)}-\Omegabhat)$ is obtained by applying the Taylor 
expansion of the function  $\frac{d f(\Sb,\Omegab)}{d\Omegab}^{\top}$ around $(\Sb,\Omegabhat)$ in the point $(\Sb^{(-k)},\Omegabhat^{(-k)})$. We expand the transposed term because we consider vectors as columns. 
$$\ob_{p^2}=\left\{\frac{d f(\Sb^{(-k)},\Omegabhat^{(-k)})}{d\Omegab}\right\}^{\top}\approx \left\{\frac{d f(\Sb,\Omegabhat)}{d\Omegab}\right\}^{\top}+\frac{d^2 f(\Sb,\Omegabhat)}{d\Omegab^2}\vect(\Omegabhat^{(-k)}-\Omegabhat)+\frac{d^2 f(\Sb,\Omegabhat)}{d\Omegab d\Sb}\vect(\Sb^{(-k)}-\Sb),$$
where $\ob_{p^2}$ is the column vector of zeros of dimension $p^2$. From here it follows that
$$\vect(\Omegabhat^{(-k)}-\Omegabhat)\approx-\left\{\frac{d^2 f(\Sb,\Omegabhat)}{d\Omegab^2}\right\}^{-1}\frac{d^2 f(\Sb,\Omegabhat)}{d\Omegab d\Sb}\vect(\Sb^{(-k)}-\Sb).$$	
We have $d f(\Sb,\Omegabhat)/d\Omegab=\vect(\Omegabhat^{-1}-\Sb)$, so $d^2 f(\Sb,\Omegabhat)/d\Omegab d\Sb=-\Ib_{p^2}$, $d^2 f(\Sb,\Omegabhat)/d\Omegab^2=-\Omegabhat^{-1}\otimes\Omegabhat^{-1}$ and consequently
$$\vect(\Omegabhat^{(-k)}-\Omegabhat)\approx-(\Omegabhat\otimes\Omegabhat)\vect(\Sb^{(-k)}-\Sb).$$
It follows that the approximation of $\LOOCV$, denoted by $\KLCV$, has the form
$$
\KLCV= -\frac{1}{n}l(\Omegabhat)+\frac{1}{2n}\sum_{k=1}^n\vect(\Omegabhat^{-1}-\Sb_k)^{\top}(\Omegabhat\otimes\Omegabhat)\vect(\Sb^{(-k)}-\Sb).
$$
After simplifying the term in the sum we finally obtain 
\begin{equation}
\label{klcvmle} 
\KLCV=-\frac{1}{n}l(\Omegabhat)+1/2n(n-1)\sum_{k=1}^n T_k
\end{equation}
where
$$T_k=\vect(\Omegabhat^{-1}-\Sb_k)^{\top}(\Omegabhat\otimes\Omegabhat)\vect(\Sb-\Sb_k).$$
This formula is equivalent to that from \eqref{bias_klcv} and we will show this in the end of the next section. Also, this formula is equivalent to the one obtained in \cite{lian2011shrinkage} who proposed it for both, MLE and MPLE.
We do not advocate using this formula for MPLE since it ignores the sparsity assumption. For this reason, we treat the case of MPLE separately in the next section. We also show that the obtained formula for the MPLE is an extension of the
formula for the MLE.

\subsection{Extension to the MPLE}
Before we propose the formula for the MPLE we formulate two auxiliary results. 
\begin{lemma}
\label{lemma1}
Let $\Ab$ and $\Omegab$ be a symmetric matrices of order $p$. The following identity holds
\begin{equation}
\label{eq:Mp}
(\Omegab\otimes\Omegab)\vect(\Ab)=\Mb_p(\Omegab\otimes\Omegab)\vect(\Ab), 
\end{equation}
where $\Mb_p=1/2(\Ib_{p^2}+\Kb_p)$, and $\Ib_{p^2}$ and $\Kb_p$ are identity matrix and commutation matrix of order $p^2$ respectively.
\end{lemma}
Commutation matrix $\Kb_p$ is a square matrix of dimension $p^2$ that has the property $\Kb_p \vect(\Ab)=\vect(\Ab)^{\top}$ for any matrix $\Ab$ of dimension $p$.
\begin{lemma}
\label{lemma2}
Let $\Ab$ be a symmetric matrix of order $p$ and $\xb,\yb$ any vectors of dimension $p$. Then the value of the bilinear form 
$$B(\xb,\yb)=\xb^{\top}\Ab\yb,$$
when $i$th row (column) of the matrix $\Ab$ is set to zero is the same as the value of $B(\xb,\yb)$ when $i$th entry of the vector $\xb$ ($\yb$) is set to zero.
\end{lemma}
The proof of Lemma \ref{lemma1} is given in the Appendix, while Lemma 2 is obtained by straightforward calculation.
Note that according to the Lemma \ref{lemma1} 
\begin{equation}
\label{Ti} 
T_k=\vect(\Omegabhat^{-1}-\Sb_k)^{\top}\Mb_p(\Omegabhat\otimes\Omegabhat)\vect(\Sb-\Sb_k),
\end{equation}
and that $2\Mb_p(\Omegabhat\otimes\Omegabhat)$ is an estimator of the asymptotic covariance matrix of $\Omegabhat$ \citep{fried2009robust}.  
\newline
To obtain the formula for the MPLE we assume standard conditions like in \cite{lam2009sparsistency} that guarantee sparsistent estimator. These conditions 
imply that $\lambda\rightarrow 0$ when $n\rightarrow \infty$, so we use formula (\ref{klcvmle}), derived for the MLE, as an approximation in the penalized case. By sparsistency, with probability one the zero 
coefficients will be estimated as zero when $n$ tends to infinity. This means that asymptotically the covariances between zero elements and nonzero elements are equal to zero. Thus, to obtain the term $T_k$ for the MPLE we do not 
only plug in the expression $\Omegabhat_{\lambda}$ in formula (\ref{Ti}), but we also set the elements of the matrix $\Mb_p(\Omegabhat_{\lambda}\otimes\Omegabhat_{\lambda})$ that correspond to covariances between zero and nonzero elements to zero.  
According to Lemma \ref{lemma2} this is equivalent to setting the corresponding entries of vectors 
$\vect(\Omegabhat_{\lambda}^{-1}-\Sb_k)$ and $\vect(\Sb^{(-k)}-\Sb)$ to zero, i.e. we define
$$T_k(\lambda)=\vect[(\Omegabhat_{\lambda}^{-1}-\Sb_k)\circ \Ib_{\lambda}]^{\top}\Mb_p(\Omegabhat_{\lambda}\otimes\Omegabhat_{\lambda})\vect[(\Sb-\Sb_k)\circ \Ib_{\lambda}],$$
where $\Ib_{\lambda}$ is the indicator matrix, whose entry is $1$ if the corresponding entry in the precision matrix $\Omegabhat_{\lambda}$ is nonzero and zero if the corresponding entry in the precision matrix is zero.
The obtained formula involves matrices of order $p^2$, which entails high cost in terms of both, memory usage and floating-point operations. For this reason, we rewrite the formula in a way that it is computationally feasible. 
Applying the Lemma \ref{lemma1} and the identity $\vect(\Ab\Bb\Cb)=(\Cb^{\top}\otimes \Ab)\vect \Bb$ we obtain 
\begin{equation}
\label{bias_klcv2}
T_k(\lambda)=\vect\{(\Omegabhat_{\lambda}^{-1}-\Sb_k)\circ \Ib_{\lambda}\}^{\top}\vect[\Omegabhat_{\lambda}\{(\Sb-\Sb_k)\circ \Ib_{\lambda}\}\Omegabhat_{\lambda}].
\end{equation}
\par
To conclude this section, we show that the derived formula for MPLE is an extension of the corresponding formula for MLE, meaning that applying the MPLE formula on the MLE yields the same result like the
corresponding MLE formula. To this aim, let $\Omegabhat$ be maximum likelihood
estimator of the precision matrix, which is the MPLE for $\lambda=0$, i.e. $\Omegabhat=\Omegabhat_{\lambda}$, for $\lambda=0$. Since with probability one all the elements of $\Omegabhat$ are nonzero it follows that $\Ib_{\lambda}$ is the matrix 
with all entries equal to one. This implies that 
in the formula \eqref{bias_klcv2} we have $(\Omegabhat_{\lambda}^{-1}-\Sb_k)\circ \Ib_{\lambda}=\Omegabhat_{\lambda}^{-1}-\Sb_k$ and $(\Sb-\Sb_k)\circ \Ib_{\lambda}=\Sb-\Sb_k$, which in turn implies $T_k(\lambda)=T_k$.

\section{Implementation}
\label{sec:implementation}

In this section we show how to implement formula \eqref{bias_klcv2} efficiently. Although the formula \eqref{bias_klcv2} involves vectorization and transpose operators, they can be avoided in the implementation.
Indeed, for any matrices $\Xb=(x_{ij})$ and $\Ybmat=(y_{ij})$  it holds $(\vect\Xb)^{\top}\vect\Ybmat=\sum_{i,j}x_{ij}y_{ij}$ so it follows that $(\vect\Xb)^{\top}\vect\Ybmat$ is just the sum of
elements of the matrix $\Xb\circ\Ybmat$, i.e. $(\vect\Xb)^{\top}\vect\Ybmat=\sum_{i,j}(\Xb\circ\Ybmat)_{ij}$. Applying this on \eqref{bias_klcv2} we obtain 
$$T_k(\lambda)=\sum_{i,j}\left((\Omegabhat_{\lambda}^{-1}-\Sb_k)\circ \Ib_{\lambda}\circ[\Omegabhat_{\lambda}\{(\Sb-\Sb_k)\circ \Ib_{\lambda}\}\Omegabhat_{\lambda}]\right)_{ij}.$$
In statistical programming language R, expression $\sum_{i,j}(\Xb\circ\Ybmat)_{ij}$ can be efficiently implemented with {\tt sum(X*Y)}.

\section{Simulation study}
\label{sec:sim_study}

\begin{figure}[ht]
\centering
\begin{minipage}[b]{0.45\linewidth}
\includegraphics[scale=0.3]{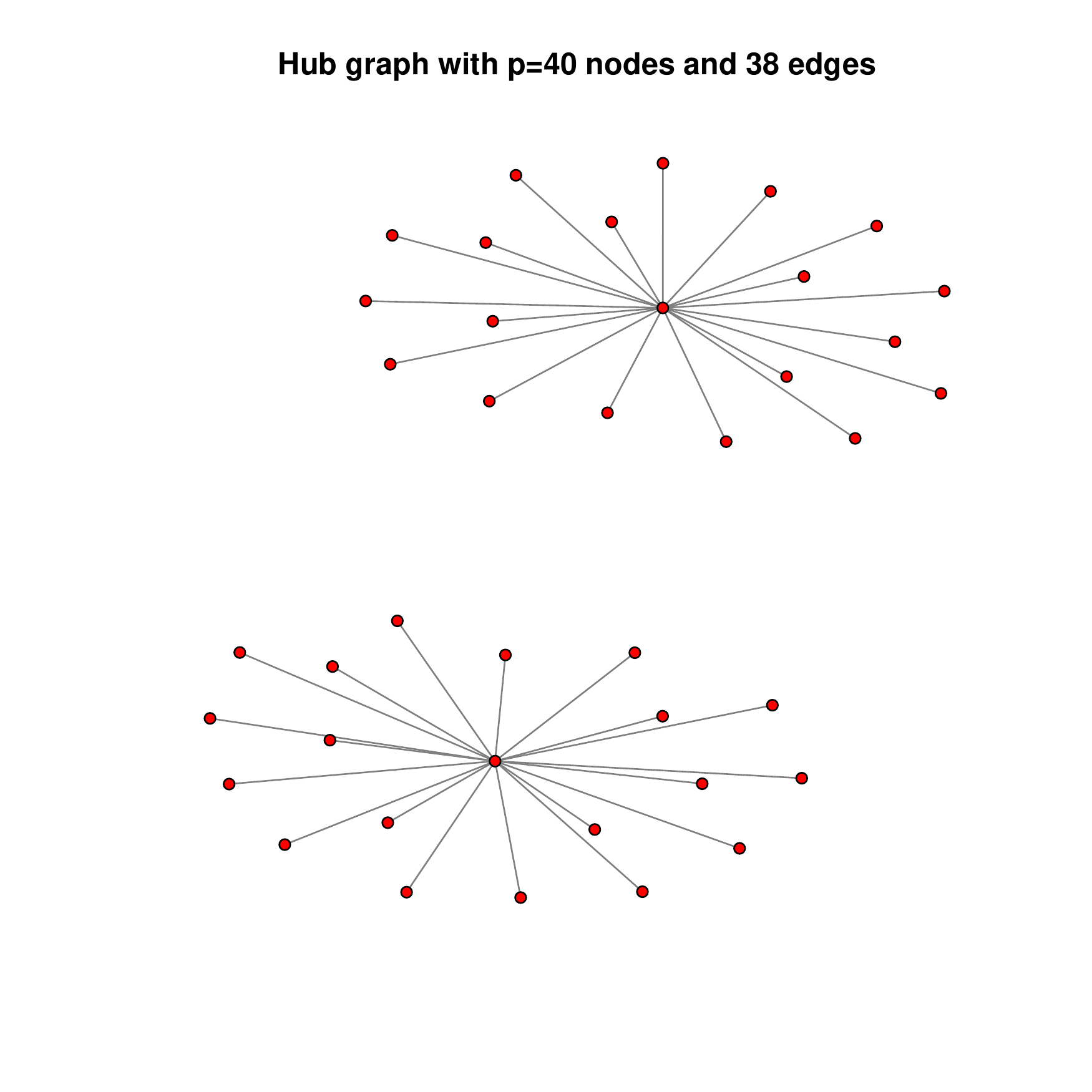}
\end{minipage}
\quad
\begin{minipage}[b]{0.45\linewidth}
\includegraphics[scale=0.3]{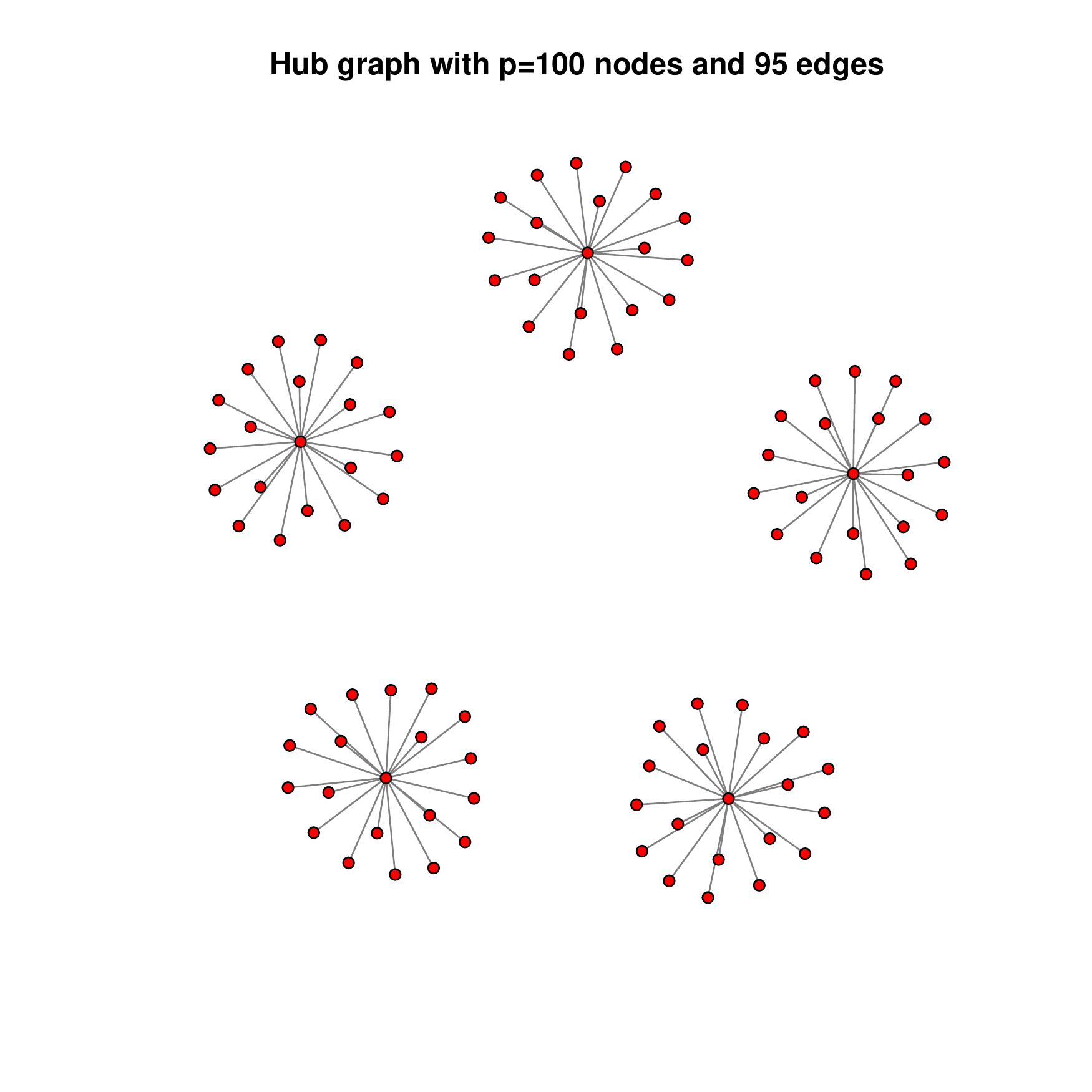}
\end{minipage}
\caption{Hub graphs with $p$=40 and $p=100$ nodes used in the simulation study.\label{fig1}}
\end{figure}

In this section we test the performance of the proposed formula in terms of Kullback-Leibler loss. We do this in case of the most popular LASSO penalty for two sparse hub graphs. The graphs have $p=40$ nodes and 38 edges and $p=100$
nodes and 95 edges. Sparsity values of these graphs are 0.049 and 0.019 respectively. The graphs are shown in figure \ref{fig1}. We omit the results for other type of graphs and for the adaptive LASSO and SCAD penalties for the same combinations of $n$ and $p$. The method was
tested for a band graph, a random graph, a cluster graph and a scale-free graph. Our estimator exhibits superior performance in all these cases.

We compare the following estimators: the KL oracle estimator, the proposed KLCV estimator, and the AIC and GACV estimators. The KL oracle estimator is that $\Omegab_{\lambda}$ in the LASSO solution path that minimizes the KL loss if 
we knew the true matrix $\Omegab$. Under each model, we generated 100 simulated data sets with different combinations of $p$ and $n$. We focus on scenario in which $n\leq p$ which is more common in applications. For the simulations
we use the {\tt huge} package in R \citep{huge}. The results are given in Tables \ref{table1} and \ref{table2}. The KLCV method is close to the KL oracle score, even for very small $n$. Overall, our method exhibits comparable 
performance to AIC and GACV in large sample size scenarios, but it clearly outperforms both when the sample size is small.  Computationally, our formula is slightly slower than the GACV since we have an additional Schur product in
the calculation of the KLCV score.
 
\begin{table}[ht]
\centering

\begin{tabular}{rrrrr}
\toprule
 p=40 & KL ORACLE & KLCV & AIC & GACV \\ 
  \hline
  n=8  & 3.68   & {\bf 3.71} & 6.46 & 26.80 \\ 
       & (0.27) & {\bf (0.28)} & (2.12) & (1.66) \\ 
  n=12 & 3.29   & {\bf 3.36 } & 6.58 & 18.34 \\ 
       & (0.26) & {\bf (0.28)} & (3.54) & (1.61) \\ 
  n=16 & 2.93   & {\bf 3.01 } & 6.62 & 13.07 \\ 
       & (0.26) & {\bf (0.26)} & (3.07) & (1.36) \\ 
  n=20 & 2.67   & {\bf 2.76 }& 6.48 & 10.08 \\ 
       & (0.23) & {\bf (0.25)} & (2.50) & (1.20) \\ 
  n=30 & 2.18   & {\bf 2.27 }& 4.59 & 5.81 \\ 
       & (0.23) & {\bf (0.25)}& (1.11) & (0.66) \\ 
  n=40 & 1.91   & {\bf 2.00 } & 3.18 & 4.13 \\ 
       & (0.19) & {\bf (0.21)} & (0.66) & (0.43) \\ 
  n=100& 1.00 & {\bf 1.04} & 1.17 & 1.32 \\ 
       & (0.10) &{\bf  (0.11) }& (0.16) & (0.14)\\ 
\bottomrule
\end{tabular}
\caption{Simulation results for hub graph with $p=40$ nodes. Performance in terms of Kullback-Leibler loss of different estimators for different sample size $n$ is showed. The results are based on 100 simulated data sets.
 Standard errors are shown in brackets. The best result is boldfaced.}
\label{table1}
\end{table}

\begin{table}[ht]
\centering
\begin{tabular}{rrrrr}
\toprule
  p=100 & KL ORACLE & KLCV & AIC & GACV \\ 
  \hline
  n=20  & 8.06 & {\bf 8.60 }& 12.24 & 28.59 \\ 
        & (0.37) &{\bf (0.45)} & (0.28) & (19.94) \\ 
  n=30  & 6.87 & {\bf 7.29} & 10.59 & 32.07 \\ 
        & (0.34) & {\bf(0.39)} & (0.41) & (2.77) \\ 
  n=40  & 5.92 & {\bf 6.34} & 9.15 & 22.48 \\ 
        & (0.30) &{\bf (0.38)} & (0.59) & (1.88) \\ 
  n=50  & 5.24 & {\bf 5.63} & 7.33 & 16.93 \\ 
        & (0.27) &{\bf (0.33)} & (0.81) & (1.40) \\ 
  n=75  & 4.08 & {\bf 4.36 }& 4.76 & 9.80 \\ 
        & (0.27) &{\bf (0.31)} & (0.71) & (0.71) \\ 
  n=100 & 3.34 & {\bf 3.57} & 3.63 & 6.81 \\ 
        & (0.19) &{\bf (0.23) }& (0.48) & (0.52) \\ 
  n=400 & 1.13 & 1.20 & {\bf 1.17} & 1.24 \\ 
        & (0.07) & (0.08) &{\bf (0.08)} & (0.07 )\\ 
  \bottomrule
\end{tabular}
\caption{Simulation results for hub graph with $p=40$ nodes. Performance in terms of Kullback-Leibler loss of different estimators for different sample size $n$ is showed. The results are based on 100 simulated data sets.
 Standard errors are shown in brackets. The best result is boldfaced.}
\label{table2}
\end{table}

\section{Using KLCV for graph estimation}
\label{sec:graph_estimation}

Information criteria, such as AIC, (E)BIC, for model selection in Gaussian graphical model are based on penalizing the likelihood with a term that involves an estimator of the degrees of freedom, which is defined as  
\begin{equation}\label{eq:df}
 \df(\lambda)=\sum_{1\leq i<j\leq p}I(\omegahat_{ij,\lambda}\neq 0),
\end{equation}
where  $(\omegahat_{ij,\lambda})_{1\leq i<j\leq p}$ are the estimated parameters \citep{yuan2007model}. As we pointed out in section 2, unlike the AIC, the (E)BIC has a graph selection consistency property. However, in sparse data 
settings both the BIC and the EBIC can perform poorly. The reason is the instability of the degrees of freedom defined in (\ref{eq:df}). As \cite{li2006gradient} points out, in the high-dimensional case there is often considerable
uncertainty in the number of non-zero elements in the precision matrix.  To overcome this uncertainty, the authors propose to use the bootstrap method to determine the statistical accuracy and the importance of each non-zero 
elements identified. One can then choose only the elements with high probability of being non-zero in the precision matrix across the bootstrap samples. Here we propose an alternative, faster approach.

\begin{figure}
\begin{center}
\subfloat[GLASSO]{\includegraphics[width=6.5cm]{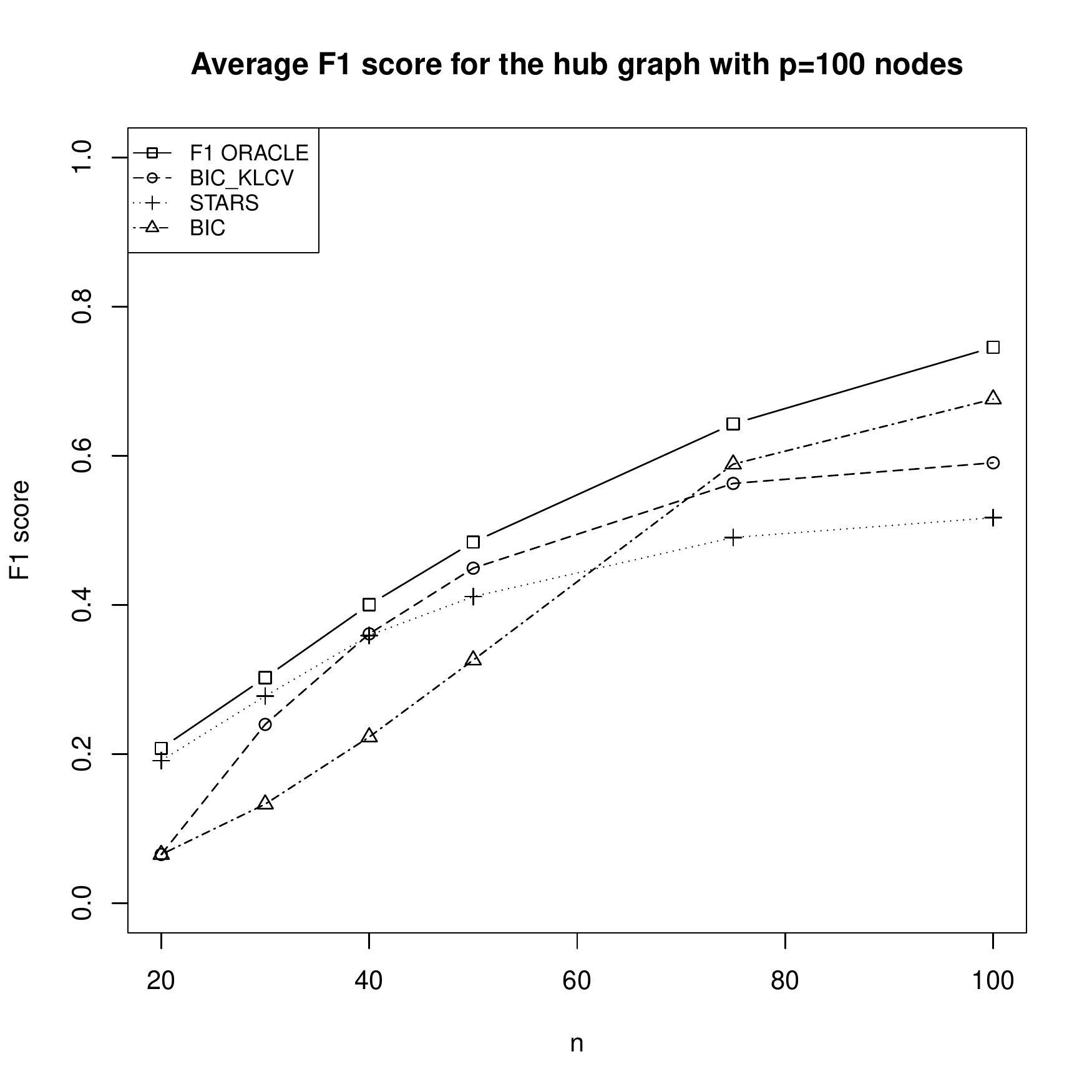}} 
\end{center}
\begin{center}
\subfloat[SCAD]{\includegraphics[width=6.5cm]{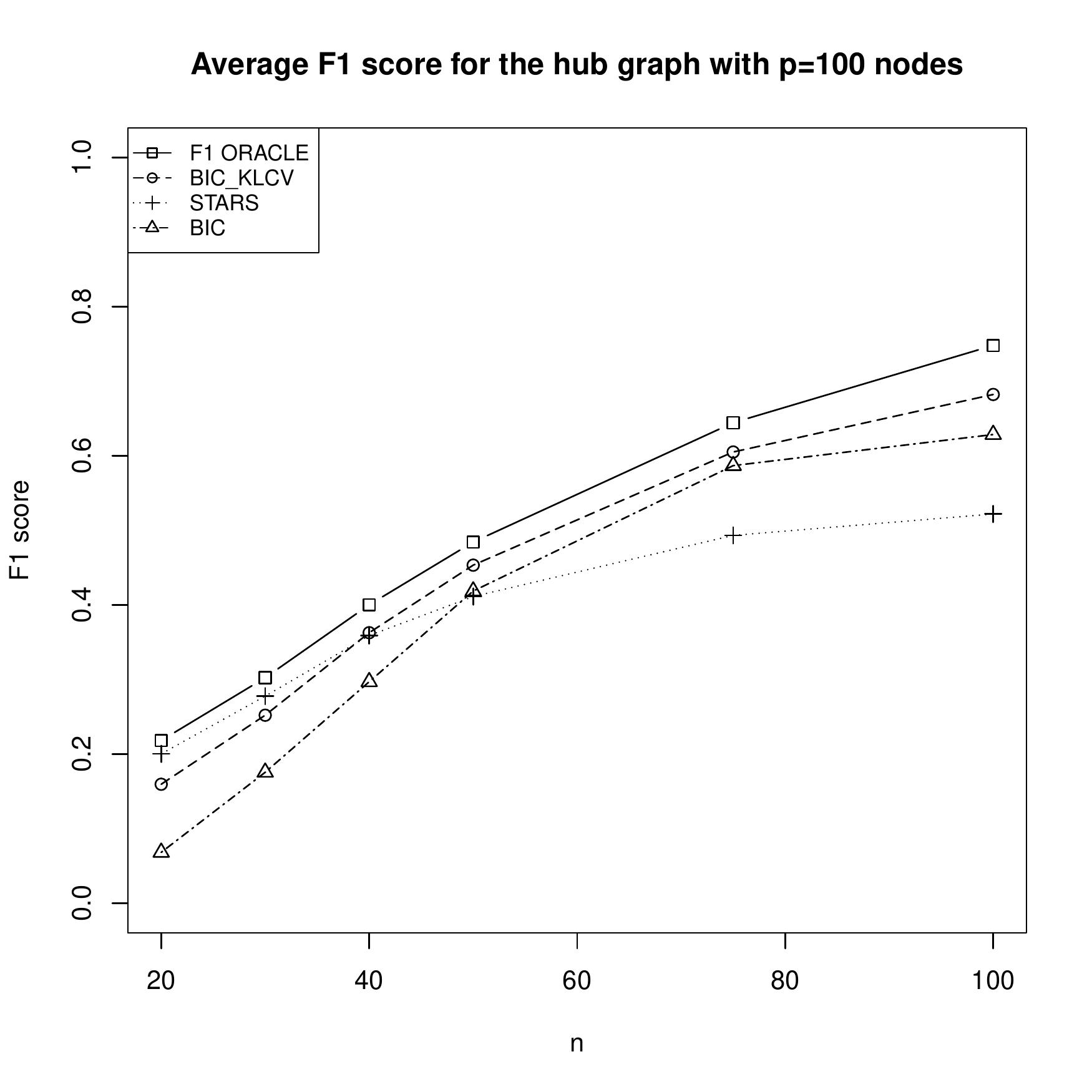}}  
\end{center}
\begin{center}
\subfloat[ADAPTIVE GLASSO]{\includegraphics[width=6.5cm]{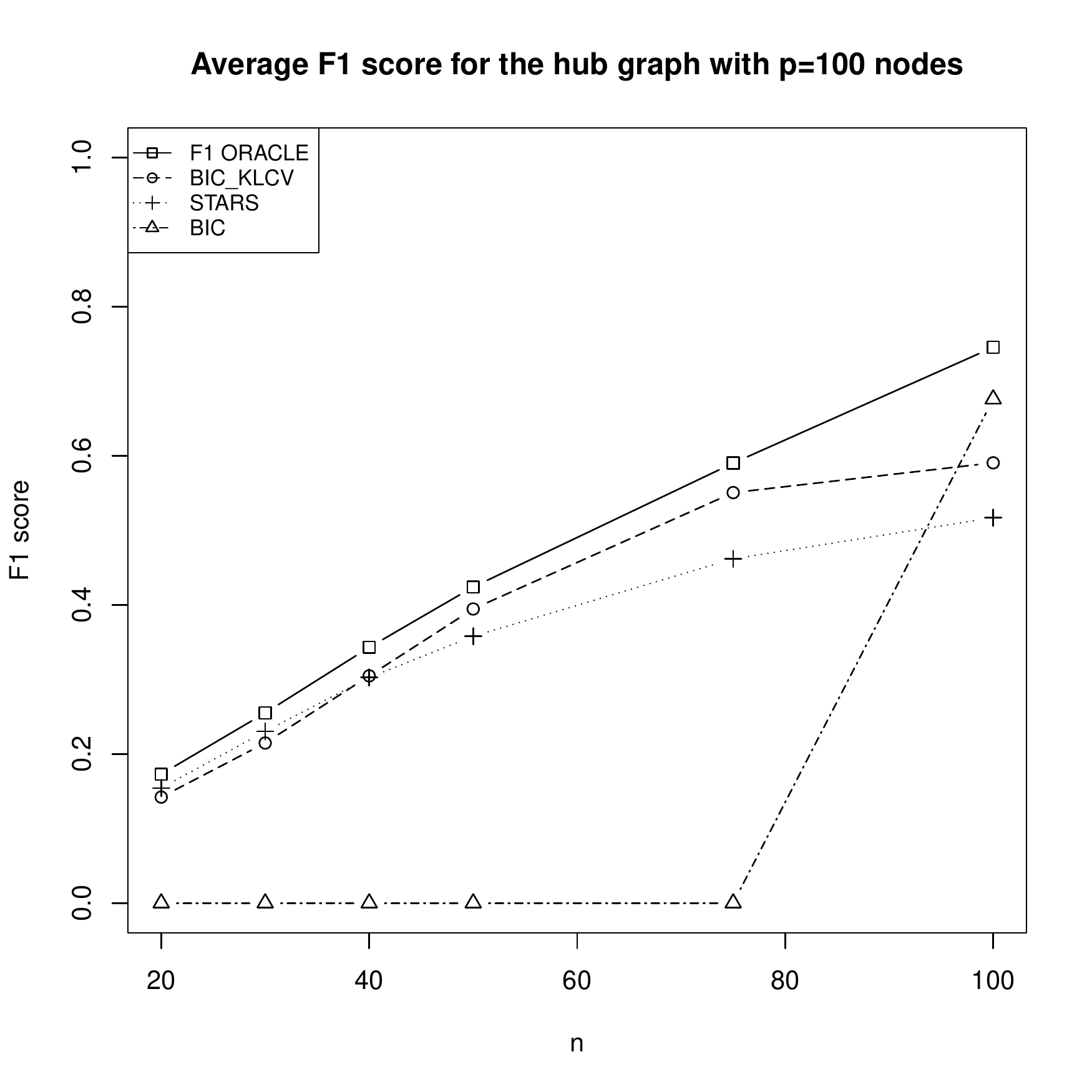}} 
\end{center}

 \caption{Simulations results for hub graph with $p=100$ nodes. Average performance in terms of F1 score of different estimators for different sample size $n$ is showed.
 The results are based on 100 simulated data sets. \label{fig2}}
\end{figure}

Recall that AIC  has the form  
$$\AIC(\lambda)=-2l(\Omegabhat_{\lambda})+2\df(\lambda),$$
where $\df(\lambda)$ is given in (\ref{eq:df}). AIC is an estimator of KL loss scaled by  $2n$. It follows that the degrees of freedom in AIC is the estimator of the bias from the KL loss scaled by $n/2$.  Since in
the proposed KL loss estimator we provide the estimator of the bias, we can use this estimator scaled by $n/2$ as the degrees of freedom in the BIC. In other words, we define the
$$\BIC_{\KLCV}(\lambda)=-2l(\Omegabhat_{\lambda})+\log n\df_{\KLCV}(\lambda),$$
where $\df(\lambda)=\frac{n}{2}\widehat{\bias}_{\KLCV}$. We compare the $\BIC_{\KLCV}$ to BIC and StARS in terms of F1 score defined as 
$$\mathrm{F}_1=\frac{2\TP}{2\TP+\FN+\FP},$$
where $\TP,\TN,\FP,\FN$ are the numbers of true positives, true negatives, false positives and false negatives. The F1 score measures the quality of a binary classifier by taking into account both true positives and negatives
\citep{baldi2000assessing,powers2011evaluation}. The larger the F1 score is, the better the classifier is. The largest possible value of the F1 score is given by the F1 oracle and is evaluated by using the true matrix $\Omegab$. 
Averaged results over 100 simulations are given in Figure \ref{fig2}. The results suggest that $\BIC_{\KLCV}$ can improve BIC for small sample sizes and can be competitive with STARS. 

\section{Conclusion}
\label{sec:conclusion}

In this article, we have proposed an alternative to cross-validation in penalized Gaussian graphical models. In simulation study we show that the estimator that we propose is the best
available non-computational method for selecting a predictively accurate model in sparse data settings for sparse Gaussian graphical models. We also illustrated that our estimator of KL loss can be useful
to for the graph selection problem.

\appendix

\section{Proof of Lemma \ref{lemma1}}
Commutation matrix $\Kb_p$ is a square matrix of dimension $p^2$ that has the property $\Kb_p \vect(\Ab)=\vect(\Ab^{\top})$. By substituting $\Mb_p=1/2(\Ib_{p^2}+\Kb_p)$ in the equality \eqref{eq:Mp} we obtain that it is equivalent to 
$$(\Omegab\otimes\Omegab)\vect(\Ab)=\Kb_p(\Omegab\otimes\Omegab)\vect(\Ab).$$
To show the above equality, we use identities  $\vect(\Ab\Bb\Cb)=(\Cb^{\top}\otimes \Ab)\vect \Bb$, $\Kb_p \vect(\Ab)=\vect(\Ab^{\top})$ and that $\Ab$ and $\Omegab$ are symmetric
$$\Kb_p\Omegab\otimes \Omegab\vect\Ab = \Kb_p\vect(\Omegab \Ab\Omegab)=\vect\{(\Omegab \Ab\Omegab)^{\top}\}=\vect(\Omegab \Ab\Omegab)=\Omegab\otimes \Omegab\vect\Ab.$$
$\Box$

\section{Calculations of the derivatives}
In the literature there are several definitions of the derivative of a function of a matrix variable. In this paper we use the definition of the derivative given in \cite{magnus2007matrix}, which is the only natural and viable 
generalization of the notion of a derivative of a vector function to a derivative of a matrix function. Let $\Fb$ be a differentiable $m\times p$
real matrix function of an $n\times q$ matrix of real variables $\Xb=(x_{ij})$. The derivative (or Jacobian matrix) of $\Fb$ at $\Xb$ is the $mp\times nq$ matrix 
$$\sD \Fb(\Xb)=\frac{\partial \vect \Fb(\Xb)}{\partial(\vect \Xb)^{\top}},$$
where the derivative of vector valued function $\fb=(f_1,\ldots,f_m)^{\top}$  of vector $\xb=(x_1,\ldots,x_n)^{\top}$ is defined as the matrix $(\partial f_i(\xb)/\partial x_j)$.
We also use the following notation for the matrix derivatives of scalar function $\phi$ of two matrix arguments, which have no common variables
\begin{align}
\label{eq:partial}
\frac{d\phi(\Xb,\Ybmat)}{d \Xb}&:=\sD_{\Xb}\phi(\Xb,\Ybmat)=\frac{\partial \phi(\Xb,\Ybmat)}{\partial (\vect \Xb)^{\top}},\\
\frac{d\phi(\Xb,\Ybmat)}{d \Xb d \Ybmat}&:=\sD_{\Xb}\left\{\sD_{\Ybmat}\phi(\Xb,\Ybmat)\right\}^{\top},
\label{eq:double_partial}
\end{align}
where $\sD_{\Xb}$ and $\sD_{\Yb}$ stress that the derivatives are with respect to $\Xb$ and $\Ybmat$, respectively.  The transpose sign of a row vector $\sD_{\Ybmat}\phi(\Xb,\Ybmat)$ in \eqref{eq:double_partial} is necessary since, in this framework, the 
calculus is developed for column vector valued functions.\par
Regarding the previous comment, in matrix calculus attention should be payed to the dimension of the matrix. Taking the derivative of the matrix is not the same as taking the derivative of the transpose matrix. Indeed, for the matrix $\Xb$
the derivative of the transpose function $\Fb(\Xb) = \Xb^{\top}$ is not an identity matrix, but it is given by 
$\sD \Fb(\Xb) = \Kb_{p}$, where  $\Kb_{p}$  is the commutation matrix of order $p^2$. 
For more on this subject see \cite{magnus2007matrix}, on which our exposition is based on and which also contains the following results that we use.
\begin{lemma}
\label{eq:diff_results}
Let $\Xb$ be a square matrix of order $p$, $\Ab$ be a constant matrix od order $p$ and $\Ib_{p^2}$ and $\mathbf{O}_{p^2}$ the identity and the zero matrix of order $p^2$, respectively. 
The following identities hold
\begin{align}
\label{eq3:diff_det}
\sD|\Xb|&=|\Xb|\{\vect(\Xb^{-1})^{\top}\}^{\top},\\
\label{eq3:diff_tr}
\sD\tr(\Ab\Xb)&=\{\vect(\Ab^{\top})\}^{\top},\\
\label{eq3:diff_vec}
\sD\vect(\Xb)&=\Ib_{p^2},\\
\label{eq3:diff_inv}
\sD\Xb^{-1}&=-(\Xb^{\top})^{-1}\otimes \Xb^{-1},\\ 
\label{eq3:diff_const}
\sD\Ab&=\mathbf{O}_{p^2}. 
\end{align}
\end{lemma}
For the derivation of the  KLCV we need to show the following equalities
\begin{align}
\label{sec3:derivatives_kl1}
\frac{d f(\Sb,\Omegab)}{d\Omegab}&=\vect(\Omegab^{-1}-\Sb)^{\top},\\
\label{sec3:derivatives_kl2}
\frac{d^2 f(\Sb,\Omegab)}{d\Omegab d\Sb}&=-\Ib_{p^2},\\
\label{sec3:derivatives_kl3}
\frac{d^2  f(\Sb,\Omegab)}{d\Omegab^2}&=-\Omegab^{-1}\otimes\Omegab^{-1}.
\end{align}

We establish \eqref{sec3:derivatives_kl1} by using formulas for the derivatives of the determinant and the trace \eqref{eq3:diff_det} and \eqref{eq3:diff_tr}, the chain rule and the fact that matrices $\Omegab$ and $\Sb_k$ are symmetric. Equality 
\eqref{sec3:derivatives_kl2} follows from \eqref{sec3:derivatives_kl1}, \eqref{eq3:diff_vec} and \eqref{eq3:diff_const}. Finally, \eqref{sec3:derivatives_kl3} follows  from  \eqref{sec3:derivatives_kl1}, \eqref{eq3:diff_inv}
and \eqref{eq3:diff_const}.

\bibliographystyle{Chicago}
\bibliography{references}


\end{document}